\documentclass[runningheads]{llncs}
\usepackage{graphicx}
\usepackage{hyperref}
\begin{document}
\title{Stars, Stripes, and Silicon: Unravelling the ChatGPT's All-American, Monochrome, Cis-centric Bias}
\titlerunning{Stars, Stripes, and Silicon}
%
\author{Federico Torrielli\orcidID{0000-0001-8037-8828}}
%
\authorrunning{Torrielli}
%
\institute{University of Torino, Department of Computer Science, Torino, Italy}
%
\maketitle              
\begin{abstract}
This paper investigates the challenges associated with bias, toxicity, unreliability, and lack of robustness in large language models (LLMs) such as ChatGPT. It emphasizes that these issues primarily stem from the quality and diversity of data on which LLMs are trained, rather than the model architectures themselves. As LLMs are increasingly integrated into various real-world applications, their potential to negatively impact society by amplifying existing biases and generating harmful content becomes a pressing concern. The paper calls for interdisciplinary efforts to address these challenges. Additionally, it highlights the need for collaboration between researchers, practitioners, and stakeholders to establish governance frameworks, oversight, and accountability mechanisms to mitigate the harmful consequences of biased LLMs. By proactively addressing these challenges, the AI community can harness the enormous potential of LLMs for the betterment of society without perpetuating harmful biases or exacerbating existing inequalities.
\end{abstract}
\section{Introduction}

Bias, toxicity, unreliability and lack of robustness are interrelated issues that plague large language models (LLMs). Given that LLMs are utilised in various real-world applications, including language translation \cite{team2022NoLL,Chowdhery2022PaLMSL,Radford2019LanguageMA,Lewis2019BARTDS}, search engines \cite{perplexity_ai,metaphor}, and scientific literature summarisation \cite{Taylor2022GalacticaAL}, it is crucial that production-ready LLMs exhibit minimal bias and do not generate harmful content. However, the current state of language models faces significant challenges in this regard.

In this work, we analyse the issues that affect state-of-the-art language models such as ChatGPT, an RLHF-augmented \cite{Bai2022TrainingAH} chatbot based on GPT-3.5 \cite{brown2020language}. These problems stem less from the model architectures themselves and more from the fact that the models are trained on massive collections of uncurated data from the Internet \cite{Wei2023AnOO}. While LLMs gain much of their knowledge and capabilities from the scale of data, we argue for the use of high-quality, curated datasets over stronger content filters as a solution. We discuss why this superficial approach is problematic.

Addressing problems like bias is crucial not only to ensure theoretical soundness but also to implement practically before LLMs become integrated into daily technologies used by the general public. Their widespread adoption could amplify the negative societal effects of model flaws on vulnerable groups \cite{Zhuo2023ExploringAE}. Techniques including data curation, model interpretability, and adversarial testing can help, but fully mitigating these issues will require ongoing collaboration between researchers and practitioners. Overall, the development of fair, ethical and trustworthy AI systems must be an interdisciplinary effort prioritised in the development of advanced technologies like LLMs.
\section{The Bias Bazaar}

LLMs generate responses with a coherent and fluent natural language structure, creating an illusion of authority and credibility. This presentation format implies an intelligence that encourages users to accept the outputs at face value, exacerbating the human tendency to trust autonomous systems that reduce cognitive load \cite{Talboy2023ChallengingTA}. These factors could impede the ability to distinguish facts from falsehoods and rational from irrational reasoning in LLM outputs. In this context, accepting fake news, toxic content and bias is easy and seen as normal by a inexperienced user.

It is crucial to note that LLMs like chatbots currently generate all responses as text, though recent models such as GPT-4 experiment with multi-modal approaches \cite{openai2023gpt4}. Each step towards humanising these interfaces makes it increasingly difficult to approach them with a critical perspective. Despite this, humans tend to view text as more credible and accurate than other media, given vision's dominance as a sense \cite{Wang2012VisualAI,doi:10.1146/annurev-psych-122414-033400}.

Biases represent the discrepancy between rational and heuristic behaviour \cite{Tversky1978JudgmentUU,Ahmad2017InstitutionalIB}. As of 2023, cognitive science has identified innumerable cognitive biases \cite{Basta2019EvaluatingTU,Beltagy2019SciBERTAP,Kurita2019MeasuringBI,Sheng2019TheWW,Zhang2020HurtfulWQ} which can lead to flawed reasoning, irrationality, and potentially harmful consequences \cite{Talboy2023ChallengingTA}. Prominent biases studied in the literature include cultural, gender, nationality, political, and ethical biases.

While recent work has examined the presence of specific biases in various models \cite{Talboy2023ChallengingTA}, the scale and complexity of LLMs today make comprehensive auditing and remediation challenging. As models continue to increase in capability and adoption, governance frameworks, oversight, and accountability are urgently needed. Reliance on biased algorithms and data can directly and negatively impact marginalised groups through unfair treatment, discrimination, or by influencing consequential decisions.
\section{ChatGPT Waves the American Flag}

Recent progress in LM design has led to increasingly large models that demonstrate strong capabilities in various natural language tasks \cite{Bender2021OnTD}. However, larger models also introduce and amplify biases present in their training data \cite{Zhao2023ASO,Bender2021OnTD}. For models trained primarily on raw data ingestion, the characteristics of the training data have significant influence on model performance and biases.

For example, contemporary language models like GPT are trained on datasets comprised primarily of American English data \cite{Luo2023APM}. This restricted data source limits the diversity of perspectives and linguistic knowledge that can be acquired by the model. As language models become more capable and ubiquitous, it is crucial to consider the ramifications of biases that can be perpetuated and even exacerbated in these systems. Addressing this issue will require developing methods for building models that learn from diverse, high-quality data as well as techniques for identifying and mitigating biases. Overall, the capabilities and limitations of large language models depend strongly on the data used to train them. For a model, language bias is statistical-sampling bias, and ultimately the latter is knowledge bias \cite{Luo2023APM}.

A naive approach to addressing lack of diversity in datasets would be to simply increase the size of the training set, with the expectation that a larger dataset would inherently capture greater diversity. However, as noted by Bender et al. \cite{Bender2021OnTD}, increased size alone does not necessarily guarantee increased diversity. The underlying motivation here is that perspective is linked to language itself (as exemplified by the 'elephant' example discussed in \cite{Luo2023APM}), and language represents a powerful and distinguishing feature in how information is filtered and conveyed.

\section{The Larger They Get, the Larger Their Shadow is}


The representation of viewpoints in large language models (LLMs) is primarily governed by frequency, which is an inherent aspect of their architecture and not easily modifiable \cite{Mills2019DisabilityBA}. Biases often originate from extensive, unfiltered corpora and can persist even when safety filters are employed in the architecture \cite{openai2023gpt4}. The dominance of frequency in LLMs gives rise to a critical issue: the under-representation of less frequent data. Consequently, majority viewpoints tend to overshadow minority perspectives.

For instance, Wikipedia is frequently among the most representative sources in corpora used for LLM training. However, its content is predominantly authored by males, with female contributors constituting less than 15\% of the total \cite{barera2020mind}. Furthermore, training datasets such as CommonCrawl\footnote{https://commoncrawl.org/the-data/} and The Pile \cite{gao2020pile} have been found to contain high levels of toxic, racist, or sexist content \cite{Gehman2020RealToxicityPromptsEN}. 
This underscores the significance of carefully selecting and curating training data to mitigate biases in LLMs.

The widespread adoption of the "\textit{bigger is better}" paradigm in the context of large language models presents both ethical and computational challenges. While it is evident that larger training datasets yield improved performance for LLMs, the necessity of human involvement in dataset curation or generation cannot be ignored. This labour is frequently carried out by crowdworkers who receive inadequate compensation, lack essential protections, and are exposed to harmful content throughout their workday \cite{Gray2019GhostWH,Hara2017ADA,Kittur2013TheFO,Kneese2014UnderstandingFL}. Moreover, due to the sheer size and continuous growth of these datasets, assessing their quality in terms of bias identification and toxicity presence becomes increasingly difficult.

An additional concern arises from the primary source of LLM data: private corporations. The majority of LLM research, models, and datasets are developed by these entities, as LLM-related processes are resource-intensive and consequently, amplify corporate influence \cite{Stallman2009FreeSF}. This dynamic generates significant implications for the scientific community. Corporations prioritise product development over research, leading to the prevalence of proprietary technology, which inherently obstructs free access to the underlying research and perpetuates their dominance in the field \cite{Goetze2021BiggerIB}.

\section{Complete the Sentence: All you Need is... [Violence]}


The current trend in LLMs for secure human-chatbot interactions involves reinforcement learning with human feedback (RLHF) \cite{Griffith2013PolicySI,christiano2023deep} and standard safety mechanisms. RLHF ensures safety in typical "naive" interactions \cite{Zhuo2023ExploringAE,Greshake2023MoreTY}, while safety mechanisms have been found to be less reliable against prompt injection \cite{Greshake2023MoreTY}. A majority of prompt injection techniques utilise storytelling, an effective method capable of diverting LLMs like ChatGPT from generating innocuous content and instead producing harmful narratives. The underlying motivations behind this phenomenon remain unclear; however, some attribute it to a semiotic-simulation theory known as "\textit{The Waluigi Effect}"\footnote{\href{https://www.lesswrong.com/posts/D7PumeYTDPfBTp3i7/the-waluigi-effect-mega-post}{https://www.lesswrong.com/posts/D7PumeYTDPfBTp3i7/the-waluigi-effect-mega-post}}, wherein the creation of an ideal simulation environment allows the LLM greater freedom to improvise. As larger and more sophisticated models increasingly exhibit a tendency to reproduce human common misconceptions \cite{Lin2021TruthfulQAMH}, it is anticipated that this issue will continue to exacerbate unless more effective countermeasures are developed and implemented.

Revising the harmful content in primary training datasets is the most effective approach for mitigating the generation of toxic output from large language models. It is widely acknowledged that amidst the vast data sources used for training GPT-x models, harmful and toxic content can be found; these models' initial datasets encompass data from unreliable news sites as well as quarantined and banned subreddits \cite{Gehman2020RealToxicityPromptsEN}. Even when present in smaller quantities, such data has been shown to be more salient for the model and considerably more challenging to "unlearn" \cite{Koh2017UnderstandingBP,Carlini2018TheSS}.

With minimal or no prompting, models have been observed to generate potent and offensive content targeting minority groups and LGBTQIA+ individuals \cite{Nozza2022MeasuringHS}, thereby supporting the aforementioned hypothesis.

\section{All Bias is Language Bias}

A prevalent misconception is that these models exhibit cognitive bias, akin to those found in human decision-making \cite{Talboy2023ChallengingTA}. However, it is essential to clarify that large language models do not possess cognitive bias, as they lack cognitive abilities \cite{Mahowald2023DissociatingLA}. Instead, the biases observed in these models stem from language bias inherent in the data they are trained on \cite{Luo2023APM}. Cognitive biases emerge from cognitive processes, which involve conscious and unconscious thinking, perception, memory, and problem-solving. In contrast, large language models, including ChatGPT, function as complex pattern-recognition algorithms that learn to generate text based on statistical correlations within the data they have been trained on, rather than exhibiting any form of cognition or understanding. Language bias arises from the inherent biases present in the training data, which are a reflection of human culture, values, and beliefs. Consequently, these biases can be observed in the generated text, leading to potential misunderstandings about the presence of cognitive bias. To better understand language bias in large language models, the relationship between training data and generated text must be examined. As these models learn from vast amounts of text, they inevitably acquire the biases present in those texts.

\section{Words are Powerful: Unintended Consequences of Real-World AI Misadventures}

In light of the growing concerns about bias in LLMs like ChatGPT, it is essential that these models are developed to be explainable, transparent, unbiased, fair, verifiable, and accountable for every decision \cite{Meo2022ExplainableIT,vanDis2023ChatGPTFP}. Despite these requirements, many current LLMs are deployed without fully addressing these issues, raising questions about whether our expectations are too high or if these models are not yet ready to be products. The accelerated release of unsafe models by companies may be contributing to the difficulty in refining these models to meet these criteria, necessitating further research.

Furthermore, the extended use of unsafe models in daily life has the potential to jeopardise crucial sectors such as healthcare, medicine, code safety, journalism, online content, and spam prevention, among others \cite{Harrer2023AttentionIN,Jones2022CapturingFO}. Models like GPT-4\footnote{Pre-alignment model} have been shown to be capable of creating misinformation scenarios and spreading toxic and biased content with ease \cite{openai2023gpt4}. It is imperative that the development and deployment of LLMs prioritise safety and responsibility to prevent adverse consequences in these vital areas.

\subsection{Healthcare and Medicine}

\textbf{Positive Impacts:} ChatGPT can streamline the healthcare industry by providing quick and accurate responses to common medical questions \cite{Kung2022PerformanceOC,Gilson2023HowDC}, thereby saving time for medical professionals. Additionally, it can aid in the analysis of medical records and help identify patterns or trends that might otherwise go unnoticed.

\noindent\textbf{Negative Impacts:} If ChatGPT is not properly trained or its knowledge base is outdated, it may provide incorrect or potentially harmful medical advice. This could lead to dangerous consequences for patients and healthcare providers.

\subsection{Code Safety}

\textbf{Positive Impacts:} ChatGPT can serve as an effective tool for code review \cite{He2023LargeLM}, detecting potential bugs, and suggesting improvements to existing code. This could improve overall software quality and reduce the likelihood of security vulnerabilities.

\noindent\textbf{Negative Impacts:} For now, ChatGPT generates flawed or unsafe code suggestions \cite{Khoury2023HowSI}, and it could inadvertently introduce security risks or software bugs in the future, compromising the safety and reliability of the developed software.

\subsection{Journalism}

\textbf{Positive Impacts:} ChatGPT can assist journalists in drafting articles quickly and efficiently. It can also help in the generation of news summaries, translations, and content personalisation for readers \cite{pavlik2023collaborating}.

\noindent\textbf{Negative Impacts:} The potential for ChatGPT to generate biased or misleading content poses a risk to journalistic integrity \cite{pavlik2023collaborating,rudolph2023chatgpt,leiter2023chatgpt}. If unchecked, it could contribute to the spread of misinformation and undermine public trust in news sources.

\subsection{Online Content}

\textbf{Positive Impacts:} ChatGPT can help generate engaging and relevant content for websites, blogs, and social media platforms, assisting content creators and marketers in their efforts to reach and captivate audiences.

\noindent\textbf{Negative Impacts:} The ease with which ChatGPT can generate content may result in an oversaturation of low-quality or misleading information online. Additionally, it could be used to create and spread fake news, deepfakes, and other manipulative content \cite{rudolph2023chatgpt,vanDis2023ChatGPTFP,alkaissi2023artificial}.

\subsection{Spam Prevention}

\textbf{Positive Impacts:} ChatGPT can be utilised to develop advanced spam filters, capable of identifying and blocking spam messages more effectively by understanding the semantic meaning of text, rather than relying solely on keywords or patterns.

\noindent\textbf{Negative Impacts:} Conversely, ChatGPT can also be employed by malicious actors to generate sophisticated spam messages that bypass existing filters \cite{mansfield2023weaponising,borji2023categorical}, leading to an increase in unwanted and potentially harmful content in users' inboxes.

\section{Conclusion: the Avalanche Effect}

In conclusion, the concerns surrounding bias, toxicity, unreliability, and lack of robustness in large language models (LLMs) such as ChatGPT are multifaceted and significant. This paper highlights that the primary challenge lies in the diversity of data on which LLMs are trained. As these models become increasingly integrated into daily technologies and real-world applications, their potential to negatively impact society by amplifying existing biases and generating harmful content is intensified.

One particularly noteworthy concern we should raise is the possible \textit{"avalanche effect"}, wherein future LLMs could inadvertently include generated content from previous LLMs in their training data. This effect could result in a self-perpetuating loop of biased and potentially harmful content being propagated across generations of models, exacerbating the issues that already plague these systems. Consequently, it is crucial for researchers and practitioners to develop methods for mitigating these problems and ensuring the development of fair, ethical, and trustworthy AI systems.

\bibliographystyle{custom}
\bibliography{custom}

\begin{thebibliography}{10}
\providecommand{\url}[1]{\texttt{#1}}
\providecommand{\urlprefix}{URL }
\providecommand{\doi}[1]{https://doi.org/#1}

\bibitem{metaphor}
\url{https://metaphor.systems/}

\bibitem{perplexity_ai}
Perplexity ai, \url{https://www.perplexity.ai/}

\bibitem{Ahmad2017InstitutionalIB}
Ahmad, Z., Ibrahim, H., Tuyon, J.: Institutional investor behavioral biases: syntheses of theory and evidence. Management Research Review  \textbf{40},  578--603 (2017)

\bibitem{alkaissi2023artificial}
Alkaissi, H., McFarlane, S.I.: Artificial hallucinations in chatgpt: implications in scientific writing. Cureus  \textbf{15}(2) (2023)

\bibitem{Bai2022TrainingAH}
Bai, Y., Jones, A., Ndousse, K., Askell, A., Chen, A., et~al.: Training a helpful and harmless assistant with reinforcement learning from human feedback. ArXiv  \textbf{abs/2204.05862} (2022)

\bibitem{barera2020mind}
Barera, M.: Mind the gap: Addressing structural equity and inclusion on wikipedia  (2020)

\bibitem{Basta2019EvaluatingTU}
Basta, C., Costa-juss{\`a}, M.R., Casas, N.: Evaluating the underlying gender bias in contextualized word embeddings. ArXiv  \textbf{abs/1904.08783} (2019)

\bibitem{Beltagy2019SciBERTAP}
Beltagy, I., Lo, K., Cohan, A.: Scibert: A pretrained language model for scientific text. In: Conference on Empirical Methods in Natural Language Processing (2019)

\bibitem{Bender2021OnTD}
Bender, E.M., Gebru, T., McMillan-Major, A., Shmitchell, S.: On the dangers of stochastic parrots: Can language models be too big? Proceedings of the 2021 ACM Conference on Fairness, Accountability, and Transparency  (2021)

\bibitem{borji2023categorical}
Borji, A.: A categorical archive of chatgpt failures. arXiv preprint arXiv:2302.03494  (2023)

\bibitem{brown2020language}
Brown, T.B., Mann, B., Ryder, N., Subbiah, M., Kaplan, J., et~al.: Language models are few-shot learners (2020)

\bibitem{Carlini2018TheSS}
Carlini, N., Liu, C., Erlingsson, {\'U}., Kos, J., Song, D.X.: The secret sharer: Evaluating and testing unintended memorization in neural networks. In: USENIX Security Symposium (2018)

\bibitem{Chowdhery2022PaLMSL}
Chowdhery, A., Narang, S., Devlin, J., Bosma, M., Mishra, G., et~al.: Palm: Scaling language modeling with pathways. ArXiv  \textbf{abs/2204.02311} (2022)

\bibitem{christiano2023deep}
Christiano, P., Leike, J., Brown, T.B., Martic, M., Legg, S., et~al.: Deep reinforcement learning from human preferences (2023)

\bibitem{vanDis2023ChatGPTFP}
van Dis, E.A.M., Bollen, J., Zuidema, W., van Rooij, R., Bockting, C.L.H.: Chatgpt: five priorities for research. Nature  \textbf{614},  224--226 (2023)

\bibitem{gao2020pile}
Gao, L., Biderman, S., Black, S., Golding, L., Hoppe, T., et~al.: The pile: An 800gb dataset of diverse text for language modeling (2020)

\bibitem{Gehman2020RealToxicityPromptsEN}
Gehman, S., Gururangan, S., Sap, M., Choi, Y., Smith, N.A.: Realtoxicityprompts: Evaluating neural toxic degeneration in language models. In: Findings (2020)

\bibitem{Gilson2023HowDC}
Gilson, A., Safranek, C.W., Huang, T., Socrates, V., Chi, L., et~al.: How does chatgpt perform on the united states medical licensing examination? the implications of large language models for medical education and knowledge assessment. JMIR Medical Education  \textbf{9} (2023)

\bibitem{Goetze2021BiggerIB}
Goetze, T.S., Abramson, D.: Bigger isn’t better: The ethical and scientific vices of extra-large datasets in language models. 13th ACM Web Science Conference 2021  (2021)

\bibitem{Gray2019GhostWH}
Gray, M.L., Suri, S.: Ghost work: How to stop silicon valley from building a new global underclass (2019)

\bibitem{Greshake2023MoreTY}
Greshake, K., Abdelnabi, S., Mishra, S., Endres, C., Holz, T., et~al.: More than you've asked for: A comprehensive analysis of novel prompt injection threats to application-integrated large language models. ArXiv  \textbf{abs/2302.12173} (2023)

\bibitem{Griffith2013PolicySI}
Griffith, S., Subramanian, K., Scholz, J., Isbell, C.L., Thomaz, A.L.: Policy shaping: Integrating human feedback with reinforcement learning. In: NIPS (2013)

\bibitem{Hara2017ADA}
Hara, K., Adams, A., Milland, K., Savage, S., Callison-Burch, C., et~al.: A data-driven analysis of workers' earnings on amazon mechanical turk. Proceedings of the 2018 CHI Conference on Human Factors in Computing Systems  (2017)

\bibitem{Harrer2023AttentionIN}
Harrer, S.: Attention is not all you need: the complicated case of ethically using large language models in healthcare and medicine. eBioMedicine  \textbf{90} (2023)

\bibitem{He2023LargeLM}
He, J., Vechev, M.T.: Large language models for code: Security hardening and adversarial testing (2023)

\bibitem{Jones2022CapturingFO}
Jones, E., Steinhardt, J.: Capturing failures of large language models via human cognitive biases. ArXiv  \textbf{abs/2202.12299} (2022)

\bibitem{Khoury2023HowSI}
Khoury, R., Avila, A.R., Brunelle, J., Camara, B.M.: How secure is code generated by chatgpt? ArXiv  \textbf{abs/2304.09655} (2023)

\bibitem{Kittur2013TheFO}
Kittur, A., Nickerson, J.V., Bernstein, M.S., Gerber, E., Shaw, A., et~al.: The future of crowd work. Proceedings of the 2013 conference on Computer supported cooperative work  (2013)

\bibitem{Kneese2014UnderstandingFL}
Kneese, T., Rosenblat, A., Boyd, D.: Understanding fair labor practices in a networked age (2014)

\bibitem{Koh2017UnderstandingBP}
Koh, P.W., Liang, P.: Understanding black-box predictions via influence functions. ArXiv  \textbf{abs/1703.04730} (2017)

\bibitem{Kung2022PerformanceOC}
Kung, T.H., Cheatham, M., Medenilla, A., Sillos, C., Leon, L.D., et~al.: Performance of chatgpt on usmle: Potential for ai-assisted medical education using large language models. PLOS Digital Health  \textbf{2} (2022)

\bibitem{Kurita2019MeasuringBI}
Kurita, K., Vyas, N., Pareek, A., Black, A.W., Tsvetkov, Y.: Measuring bias in contextualized word representations. ArXiv  \textbf{abs/1906.07337} (2019)

\bibitem{leiter2023chatgpt}
Leiter, C., Zhang, R., Chen, Y., Belouadi, J., Larionov, D., et~al.: Chatgpt: A meta-analysis after 2.5 months. arXiv preprint arXiv:2302.13795  (2023)

\bibitem{Lewis2019BARTDS}
Lewis, M., Liu, Y., Goyal, N., Ghazvininejad, M., Mohamed, A., et~al.: Bart: Denoising sequence-to-sequence pre-training for natural language generation, translation, and comprehension. In: Annual Meeting of the Association for Computational Linguistics (2019)

\bibitem{Lin2021TruthfulQAMH}
Lin, S.C., Hilton, J., Evans, O.: Truthfulqa: Measuring how models mimic human falsehoods. In: Annual Meeting of the Association for Computational Linguistics (2021)

\bibitem{Luo2023APM}
Luo, Q., Puett, M.J., Smith, M.D.: A perspectival mirror of the elephant: Investigating language bias on google, chatgpt, wikipedia, and youtube. ArXiv  \textbf{abs/2303.16281} (2023)

\bibitem{Mahowald2023DissociatingLA}
Mahowald, K., Ivanova, A.A., Blank, I.A., Kanwisher, N.G., Tenenbaum, J.B., et~al.: Dissociating language and thought in large language models: a cognitive perspective. ArXiv  \textbf{abs/2301.06627} (2023)

\bibitem{mansfield2023weaponising}
Mansfield-Devine, S.: Weaponising chatgpt. Network Security  \textbf{2023}(4) (2023)

\bibitem{Meo2022ExplainableIT}
Meo, R., Nai, R., Sulis, E.: Explainable, interpretable, trustworthy, responsible, ethical, fair, verifiable ai... what's next? In: Symposium on Advances in Databases and Information Systems (2022)

\bibitem{Mills2019DisabilityBA}
Mills, M., Whittaker, M.: Disability, bias, and ai (2019)

\bibitem{doi:10.1146/annurev-psych-122414-033400}
Moore, T., Zirnsak, M.: Neural mechanisms of selective visual attention. Annual Review of Psychology  \textbf{68}(1),  47--72 (2017). \doi{10.1146/annurev-psych-122414-033400}, \url{https://doi.org/10.1146/annurev-psych-122414-033400}, pMID: 28051934

\bibitem{Nozza2022MeasuringHS}
Nozza, D., Bianchi, F., Lauscher, A., Hovy, D.: Measuring harmful sentence completion in language models for lgbtqia+ individuals. In: LTEDI (2022)

\bibitem{openai2023gpt4}
OpenAI: Gpt-4 technical report (2023)

\bibitem{pavlik2023collaborating}
Pavlik, J.V.: Collaborating with chatgpt: Considering the implications of generative artificial intelligence for journalism and media education. Journalism \& Mass Communication Educator p. 10776958221149577 (2023)

\bibitem{Radford2019LanguageMA}
Radford, A., Wu, J., Child, R., Luan, D., Amodei, D., et~al.: Language models are unsupervised multitask learners (2019)

\bibitem{rudolph2023chatgpt}
Rudolph, J., Tan, S., Tan, S.: Chatgpt: Bullshit spewer or the end of traditional assessments in higher education? Journal of Applied Learning and Teaching  \textbf{6}(1) (2023)

\bibitem{Sheng2019TheWW}
Sheng, E., Chang, K.W., Natarajan, P., Peng, N.: The woman worked as a babysitter: On biases in language generation. ArXiv  \textbf{abs/1909.01326} (2019)

\bibitem{Stallman2009FreeSF}
Stallman, R.M.: Free software, free society: Selected essays of richard m. stallman (2009)

\bibitem{Talboy2023ChallengingTA}
Talboy, A.N., Fuller, E.: Challenging the appearance of machine intelligence: Cognitive bias in llms. ArXiv  \textbf{abs/2304.01358} (2023)

\bibitem{Taylor2022GalacticaAL}
Taylor, R., Kardas, M., Cucurull, G., Scialom, T., Hartshorn, A.S., et~al.: Galactica: A large language model for science. ArXiv  \textbf{abs/2211.09085} (2022)

\bibitem{team2022NoLL}
team, N., Costa-juss{\`a}, M.R., Cross, J., cCelebi, O., Elbayad, M., et~al.: No language left behind: Scaling human-centered machine translation. ArXiv  \textbf{abs/2207.04672} (2022)

\bibitem{Tversky1978JudgmentUU}
Tversky, A., Kahneman, D.: Judgment under uncertainty: Heuristics and biases: Biases in judgments reveal some heuristics of thinking under uncertainty (1978)

\bibitem{Wang2012VisualAI}
Wang, H.C., Lu, S., Lim, J.H., Pomplun, M.: Visual attention is attracted by text features even in scenes without text. Cognitive Science  \textbf{34} (2012)

\bibitem{Wei2023AnOO}
Wei, C., Wang, Y.C., Wang, B., Kuo, C.C.J.: An overview on language models: Recent developments and outlook. ArXiv  \textbf{abs/2303.05759} (2023)

\bibitem{Zhang2020HurtfulWQ}
Zhang, H., Lu, A.X., Abdalla, M., McDermott, M.B.A., Ghassemi, M.: Hurtful words: quantifying biases in clinical contextual word embeddings. Proceedings of the ACM Conference on Health, Inference, and Learning  (2020)

\bibitem{Zhao2023ASO}
Zhao, W.X., Zhou, K., Li, J., Tang, T., Wang, X., et~al.: A survey of large language models. ArXiv  \textbf{abs/2303.18223} (2023)

\bibitem{Zhuo2023ExploringAE}
Zhuo, T.Y., Huang, Y., Chen, C., Xing, Z.: Exploring ai ethics of chatgpt: A diagnostic analysis. ArXiv  \textbf{abs/2301.12867} (2023)

\end{thebibliography}

\end{document}